\documentstyle[budapest,epsfig]{article}  
\frompage{000} \topage{000}                                              
\def\rightmark{Baryon Number Transfer: Linking Elementary and Complex Interactions} 
\def\leftmark{D. Varga for NA49}
 
\title{Baryon Number Transfer in Hadron+Nucleus and Nucleus+Nucleus Collisions: a Link Between Elementary and Complex Interactions} 
\authors{
{D. Varga$^1$ for the NA49 Collaboration %
}\\[2.812mm]
{\normalsize
\hspace*{-8pt}$^1$ Roland E\"otv\"os University, \\ 
H1117 Budapest, Hungary\\[0.2ex] 
}}
 
\abstract{The baryon number transfer is studied in elementary and
	complex hadronic interactions at the CERN experiment NA49 at
	the SPS, at 158 AGeV beam energy ($\sqrt{s}=17.2 GeV$). A two
	component picture is proposed, which builds up the net proton
	distribution from a target and a projectile component. Using
	pion beam, the projectile component is experimentally
	determined for p+p and p+A interactions. A similar stopping
	behaviour of the projectile component is found for p+A and A+A
	interactions. Based on these observations, the baryon transfer
	is assumed to provide a common scale of inelasticity in p+p,
	p+A and A+A interactions. A model-independent way is proposed
	to predict the pion multiplicity in A+A.}

\keyword{baryon transfer, two component picture, multiplicity} 
\PACS{13.85, 25.75}
 
\begin{document}
 
\maketitle
\setcounter{page}{1}

\section{Introduction}\label{intro}
 
	The NA49 experiment at the CERN SPS is a large acceptance
	hadron spectrometer, based on Time Projection
	Chambers. \cite{bib1}. The detector provides precision
	tracking and particle identification via the measurement of
	the ionization energy loss in the TPCs. The system is
	completed by Time of Flight detectors and calorimeters, the
	latter providing neutron detection above $x_F=0.2$ in proton
	induced reactions. A dedicated counter is designed to control
	the centrality in p+A interactions, which measures the
	multiplicity of the recoil (grey) protons.

\section{Net proton distribution in the elementary collisions}

\begin{figure}[!ht]
\begin{center}
\psfig{figure=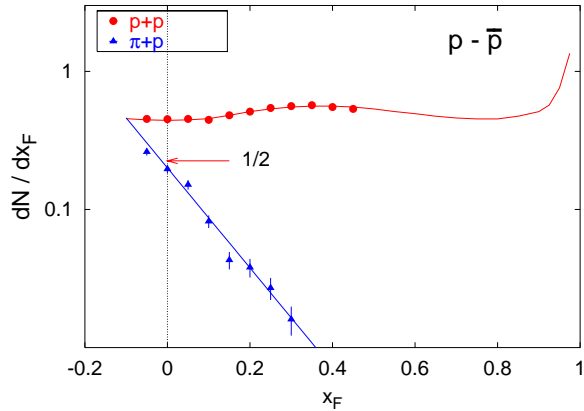,width=3.2in}
\end{center}
\vspace*{-0.6cm}
\caption{Net proton distribution in p+p and $\pi$+p interactions. The
prediction of the two component picture is that the $\pi$+p yield at
$x_F=0$ is half the p+p value 
\label{fig:pipnet}}
\end{figure}

	The net proton yield in a hadronic interaction is defined
	by $p-\bar{p}$, so antiprotons are subtracted from the protons
	to cancel the pairproduced component.  The net proton
	distribution (see Fig.\ref{fig:pipnet}) in p+p interaction is
	forward-backward symmetric, to first order flat with a
	diffractive peak. Above $x_F>0.2$ the Feynman-scaling holds,
	which predicts the invariant cross-section as a function of
	$x_F$ being independent of $\sqrt{s}$ at high
	energies. Intuitively, this symmetric distribution has two
	components: the forward region associated with the projectile,
	and the backward region with the target. In the following
	sections, the experimental study of this intuitive picture
	will be discussed.

\section{Two component picture of the elementary collisions}

\begin{figure}[!ht]
\begin{center}
\psfig{figure=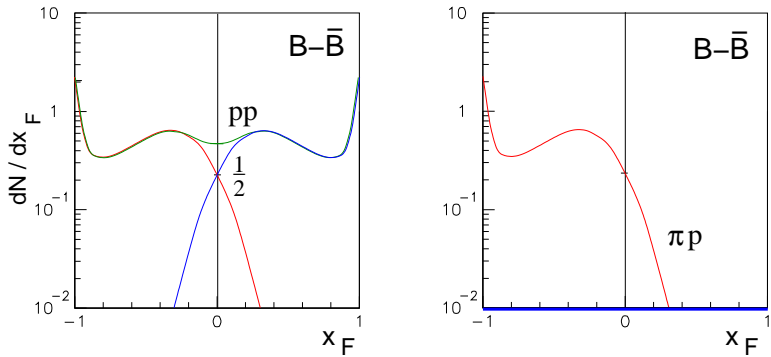,width=4.0in}
\end{center}
\vspace*{-0.6cm}
\caption{In the two component picture of p+p interactions, the net
proton yield is the sum of a target and a projectile component; in the
middle, both components contribute with half the yield (left). The
target component can be measured separately using a baryon free
projectile (right)
\label{fig:ppcart}}
\end{figure}

	The basic idea of the two component picture \cite{bib2} is
	shown on a cartoon (Fig.\ref{fig:ppcart}): the net proton
	distribution in p+p is built up from two symmetric
	components. Assuming, that \emph{the target component is
	independent of the projectile type of the interaction}, the
	two components can be experimentally separated. For the
	measurement, a baryon free projectile was used ($\pi$+p) which
	allows the measurement of the target component alone (see
	Fig.\ref{fig:ppcart}).

	As $\pi^+$ and $\pi^-$ has non-zero isospin, the produced net
	proton on the projectile side will also be non-zero (only
	$p-\bar{p} + n-\bar{n}$ would vanish). To cancel this effect,
	the average of $\pi^+ + p$ and $\pi^- +p$ collisions were used
	for the measurement. The net proton distribution in this
	averaged $\pi+p$ is presented in Fig.\ref{fig:pipnet}

\begin{figure}[!ht]
\begin{center}
\psfig{figure=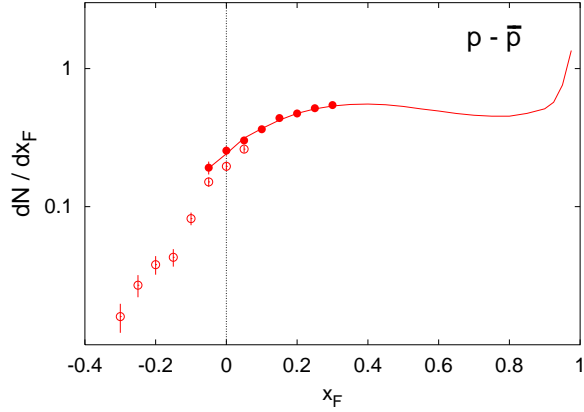,width=3.2in}
\end{center}
\vspace*{-0.6cm}
\caption{The projectile component of the net proton spectrum in p+p interaction
\label{fig:ppfwnet}}
\end{figure}

	One simple prediction of the two component picture is that the
	yield of the net protons in $\pi+p$ at $x_F=0$, so at the center
	of mass of the colliding system, has to be half of the value
	measured in $p+p$. This prediction is reasonably fulfilled. In
	the forward region, subtracting the $\pi+p$ measurement from
	the $p+p$ values, the \emph{projectile component of the net
	proton spectrum in p+p} can be obtained, shown in
	Fig.\ref{fig:ppfwnet}.
	
\section{Net proton distribution in hadron+nucleus interactions}

\begin{figure}[!ht]
\begin{center}
\psfig{figure=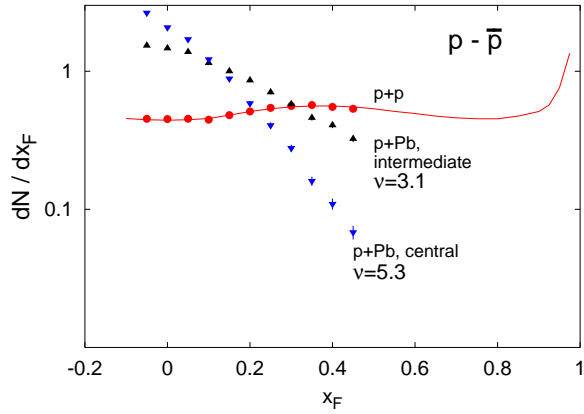,width=3.2in}
\end{center}
\vspace*{-0.6cm}
\caption{Net proton distribution in p+Pb interactions at different centralities
(different number of collisions)\label{fig:ppbnet}}
\end{figure}

	The net proton distribution in p+Pb at different centralities
	is shown in Fig \ref{fig:ppbnet}, compared to net protons in
	p+p. The centrality of the p+Pb collision is measured using a
	dedicated detector which determines the multiplicity of the
	knockout protons from the nucleus. From this multiplicity,
	applying Glauber-model calculations and detailed detector
	simulation, the number of collisions ($\nu$) is estimated. The
	distributions steepen up at higher centrality due to the
	baryon stopping phenomenon \cite{bib3}.

\section{Two component picture of hadron+nucleus interactions}

\begin{figure}[!ht]
\begin{center}
\psfig{figure=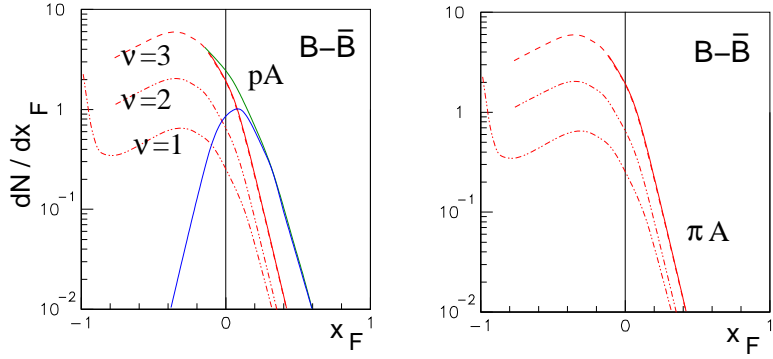,width=4.0in}
\end{center}
\vspace*{-0.6cm}
\caption{In case of p+A, the target component of the net proton yield
may have a similar shape as in p+p, but it is scaled up by the number
of collisions ($\nu$); the projectile component, as the projecile
suffered multiple collisions, gets stopped (left). The target component
can again be separately determined using pion projectile (right)
\label{fig:ppbcart}}
\end{figure}

	The proposed two component picture is more complicated
	for p+A than for p+p. The following assumption can be made: to
	first order, on the target side each collision of the
	projectile produces the target component of the net proton
	distribution in p+p, i.e. \emph{the target component in pA
	piles up in proportion with the number of collisions}. The
	projectile however, as it suffers multiple collisions,
	develops a \emph{stopped projectile component}. A cartoon
	describing the picture is shown on Fig.\ref{fig:ppbcart}.

\begin{figure}[!ht]
\begin{center}
\psfig{figure=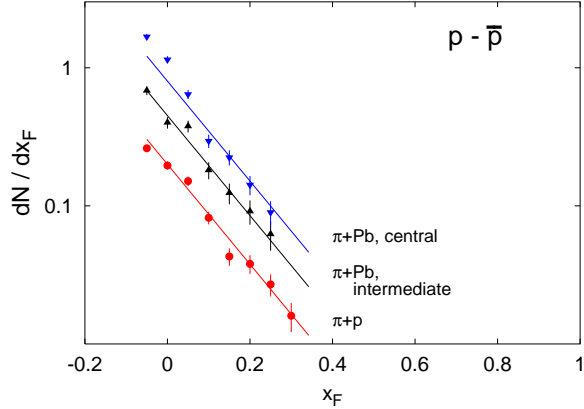,width=3.2in}
\end{center}
\vspace*{-0.6cm}
\caption{Net proton distribution in $\pi$+Pb interactions at different
centralities. Due to the pileup on the target side, the $\pi$+p yield
is scaled up in $\pi$+Pb
\label{fig:pipbnet}}
\end{figure}

	In order to determine the projectile component of the net
	proton distribution in p+A, the idea of baryon free
	projectile can be used again. According to the prediction of
	the two component picture, the target component measured by
	$\pi+A$ scales up from $\pi+p$ with the number of
	collisions; Fig.\ref{fig:pipbnet} demonstrates this effect.

\begin{figure}[!ht]
\begin{center}
\psfig{figure=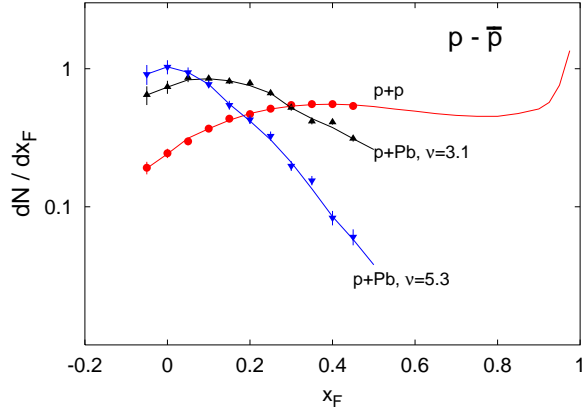,width=3.2in}
\end{center}
\vspace*{-0.6cm}
\caption{Net proton projectile component in p+Pb interaction
\label{fig:ppbfwnet}}
\end{figure}

	Subtracting the net proton distribution measured in $\pi$+Pb
	from the net protons in p+Pb at the same centrality, the
	projectile component of the net proton distribution can be
	determined in p+Pb (see Fig.\ref{fig:ppbfwnet}). A smooth
	progression of baryon number transfer is observed as a
	function of the number of collisions.

\section{Two component picture of nucleus+nucleus collisions}

\begin{figure}[!ht]
\begin{center}
\psfig{figure=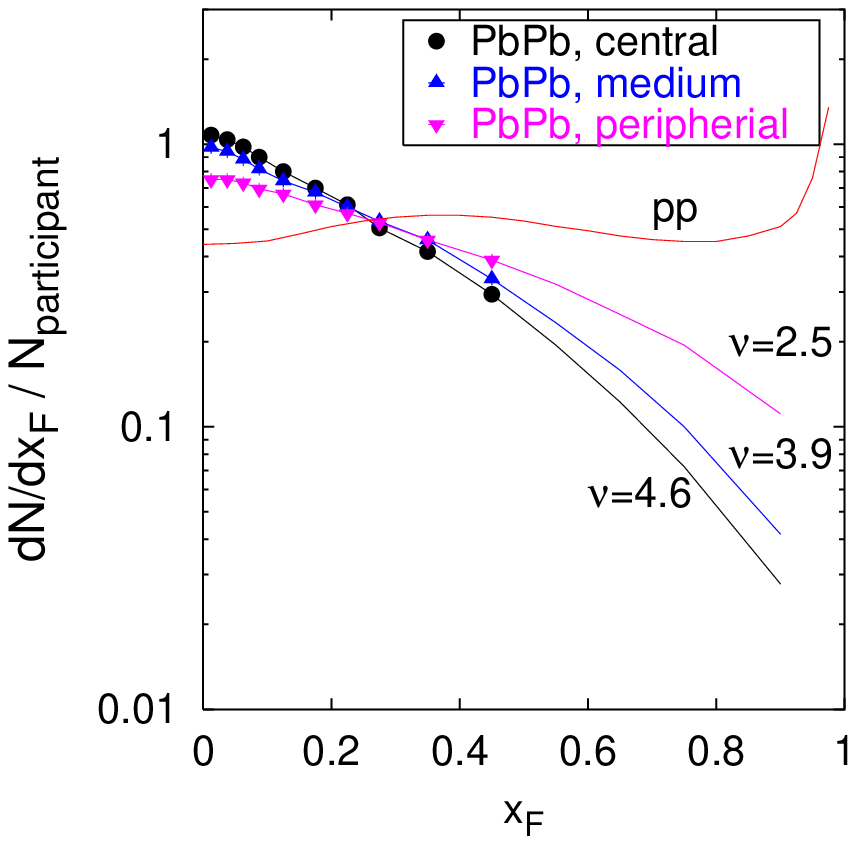,width=2.4in}
\psfig{figure=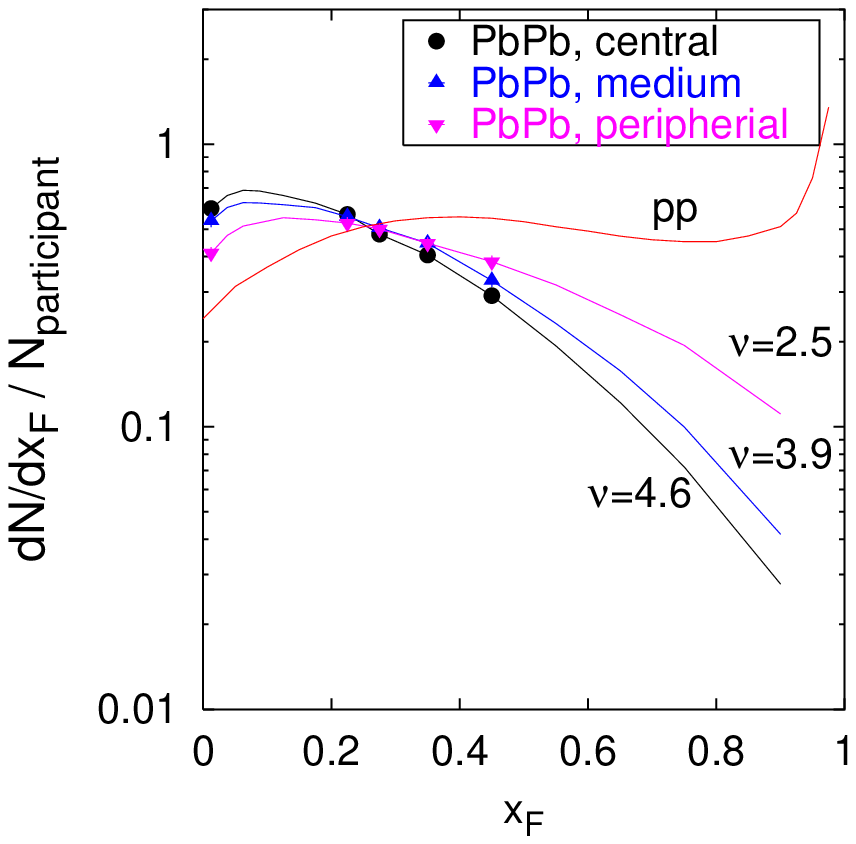,width=2.4in}
\end{center}
\vspace*{-0.6cm}
\caption{Net proton distribution in Pb+Pb interaction (left); net
proton projectile component in Pb+Pb (right)
\label{fig:pbpbnet}}
\end{figure}

	In case of AA, the baryon-free projectile concept is not
	applicable. The only constraint which provides an approximation
	of the projectile component of the net proton spectrum is the
	following: at $x_F=0$, the value of the projectile component
	is half of the measured spectra, and the target component dies
	out fast as we move forward. In Fig.\ref{fig:pbpbnet} the
	measured net proton distribution and the estimated projectile
	component is presented.

	The conclusion of the studies above is that with increasing
	centrality, the projectile component of the net proton
	distribution have a similar behaviour in p+A and A+A
	interactions, which is a smooth transition of baryon number
	transfer from the elementary to more complex interactions.

\section{Reference (isospin corrected) net proton for Pb+Pb collisions}

\begin{figure}[!ht]
\begin{center}
\psfig{figure=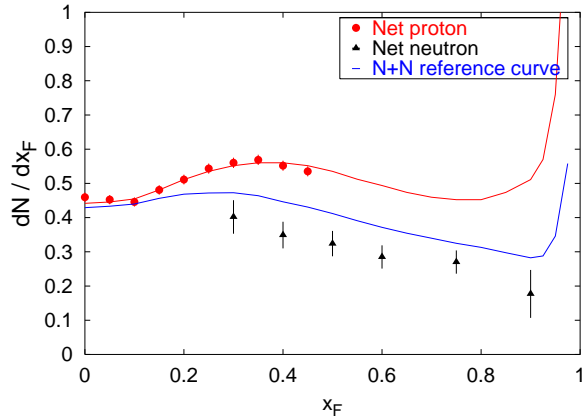,width=3.2in}
\end{center}
\vspace*{-0.6cm}
\caption{Net proton distribution in the reference nucleon-nucleon collision
\label{fig:ref}}
\end{figure}

	In the Pb nucleus, the n to p ratio is about 60 to 40, so the
	net proton distribution in a reference N+N collision has to be
	determined to take care of the neutron content; here
	N=0.4p+0.6n. Within the framework of the two component picture
	the isospin symmetry holds separately for the two hemispheres,
	so the net forward proton in n+p or n+n is equal to the net
	neutron in p+p. Relying on this assumption, the net proton
	distribution in an isospin averaged N+N collision can be
	obtained; this is shown in Fig.\ref{fig:ref}.

\section{Correlation of the pion multiplicity with the final state baryon in p+p interaction}

\begin{figure}[!ht]
\begin{center}
\psfig{figure=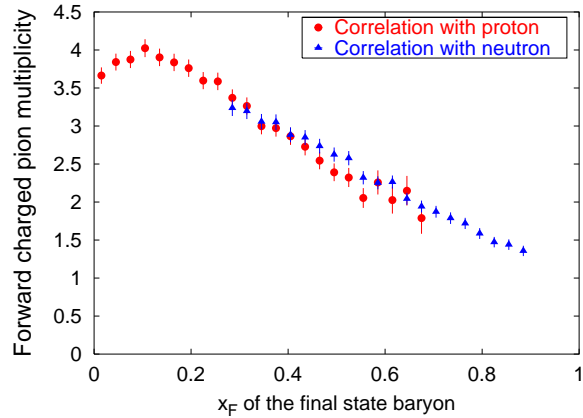,width=3.2in}
\end{center}
\vspace*{-0.6cm}
\caption{Correlation of the forward charged pion multiplicity with the
$x_F$ position of the final state baryon. Strong correlation is
observed; in those events where the final state baryon is close to the
center, the pion multiplicity increases
\label{fig:pioncorr}}
\end{figure}

	In case of the p+A and A+A collisions, the baryon number
	transfer is characterised by the centrality. In the elementary
	p+p collision the centrality can not be reliably defined, so
	the study must be done in the other way round:
	Fig.\ref{fig:pioncorr} shows that the position of the final
	state baryon characterises the pion multiplicity p+p events.\cite{bib4}

	The measurement confirms the intuitive expectation: in those
	events, where the final state baryon is fast (especially in
	the diffractive region), the multiplicity is low. If the final
	state baryon is more stopped, more pions are produced. The
	correlation is independent of the baryon isospin.

\section{Prediction of the pion multiplicity in Pb+Pb collisions from correlation measurements in p+p}

	Based on the phenomenon of baryon stopping in central p+A and
	A+A collisions, and the strong correlation of p+p observables
	with the final state baryon, the following assumption can be
	made: \emph{the baryon number transfer characterises the
	inelasticity (centrality) in p+p, p+A and A+A interactions.}
	This assumption provides a method to compare minimum bias p+p
	collisions with central A+A collisions, where the trigger bias
	is to be taken into account.

	In a p+p collision, any quantity, \mbox{e.g.} the pion
	multiplicity measured as a function of $x_F$ of the final
	state proton defines a correlation function $C(x)$. The pion
	multiplicity in minimum bias p+p then can be calculated by
	weighing $C(x)$ with the net projectile proton distribution;
	this trivial statement can be expressed in the following way:

	$$ <\pi>_{pp}=\frac{\int C(x) P_{pp} dx}{\int P_{pp}} $$
	
	where $P_{pp}$ is the net projectile proton distribution in
	p+p as a function of $x_F$. In A+A, the stopped net projectile
	proton distribution can be viewed as protons from a
	superposition of p+p (or N+N) interactions, where events with
	more stopped final state protons have larger weight; this
	means that the pion multiplicity in A+A (per participating
	nucleon) is predicted to be

	$$ <\pi>_{PbPb}=\frac{\int C(x) P_{PbPb} dx} { \int P_{PbPb} dx}$$

\begin{figure}[!ht]
\begin{center}
\psfig{figure=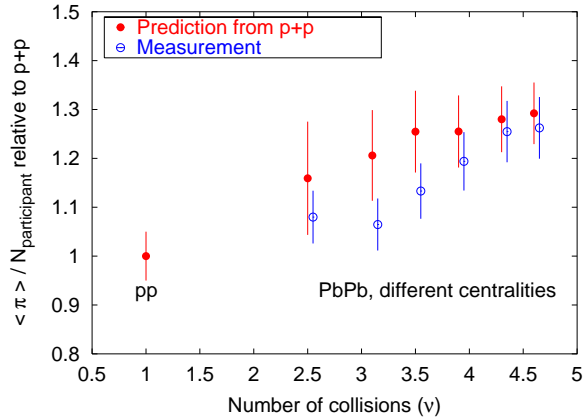,width=3.2in}
\end{center}
\vspace*{-0.6cm}
\caption{Comparison of the predicted and the measured pion
multiplicity in Pb+Pb collisions. The prediction, which assumes that
the final state baryon characterises the inelasticity also in p+p
gives a good qualitative agreement
\label{fig:pionpred}}
\end{figure}

	In this case, the net proton distribution in Pb+Pb
	($P_{PbPb}$) is to be applied. The result of this prediction
	is presented on Fig.\ref{fig:pionpred} in comparison with
	pion multiplicity measurements from PbPb interactions at
	different centralities.

	The prediction gives a very good qualitative descriprion of
	the measurements. If the assumptions are correct, this means
	that the full increase of pion multiplicity in central Pb+Pb
	interaction can be explained with a superposition of
	elementary collisions, taking proper care of the centrality
	selection.

\section{Conclusions}\label{concl}
 
	An experimental study of the two component picture of hadronic
	interactions at $\sqrt{s}$=17.2 GeV was discussed. In this
	picture, the net proton distribution is built up by a target
	and a projectile component; these components can be separated
	experimentally with a baryon free (pion) projectile. The
	projectile component of the net proton distribution in p+A and
	A+A collisions is found to behave similarly at increasing
	centrality, namely that the baryon number moves towards the
	center. Based on the assumption that the baryon transfer
	characterises the inelasticity of p+p, p+A and A+A collisions,
	a model independent procedure was developed to compare minimum
	bias p+p and central Pb+Pb collisions, which showed that the
	increase of the pion multiplicity in A+A compared to p+p
	interactions can be fully predicted only by taking care of the
	sizeable baryon stopping in A+A.

\section*{Acknowledgement}

This work was supported by the Director, Office of Energy Research,
Division of Nuclear Physics of the Office of High Energy and Nuclear
Physics of the US Department of Energy (DE-ACO3-76SFOOO98 and
DE-FG02-91ER40609), the US National Science Foundation, the
Bundesministerium fur Bildung und Forschung, Germany, the Alexander
von Humboldt Foundation, the UK Engineering and Physical Sciences
Research Council, the Polish State Committee for Scientific Research
(5 P03B 13820 and 2 P03B 02418), the Hungarian Scientific Research
Foundation (T032648, T14920, T32293 and F034707), the EC Marie Curie
Foundation, and the Polish-German Foundation.

\section*{Notes}
\begin{notes}
\item[a]
Permanent address: E\"otv\"os Lor\'and University, Department of Atomic Physics, Budapest, Hungary;\\ 
E-mail: dezso.varga@cern.ch
\end{notes}

\vfill\eject
\end{document}